
\tolerance 1500
\topskip .75truein

\font\twelverm =cmr12

\def\Fscr{{\cal F}}
\def\Oscr{{\cal O}}

\def\Tscr{{\cal T}}
\def\pq{{\vec p+\vec q}}

\def\Vscr{{\cal V}}
\def\p3{{d^3p\o {(2\pi)^3}}}
\def\pp3{{d^3p'\o {(2\pi)^3}}}
\def\e1{e_0(\vec p)}
\def\e2{e_0(\vec p+\vec q)}

\def\tt{t^{ons}}
\def\zz{\langle g\rangle}
\def\qs{\lower 5pt\hbox to 23pt{\rightarrowfill}\atop
       {s\rarrow\infty}}
\def\qq{\lower 5pt\hbox to 23pt{\rightarrowfill}\atop
      {q\rarrow\infty}}
\def\q|s|{\lower 5pt\hbox to 23pt{\rightarrowfill}\atop
        {|s|\rarrow\infty}}
\def\ton{\lower 5pt\hbox to 23pt{\rightarrowfill}\atop
       {t\rarrow \tt}}
\def\gon{\lower 5pt\hbox to 23pt{\rightarrowfill}\atop
       {g\rarrow \zz}}

\def\o{\over}
\twelverm
\overfullrule 0pt
\newdimen\offdimen
\def\offset#1#2{\offdimen #1
   \noindent \hangindent \offdimen
   \hbox to \offdimen{#2\hfil}\ignorespaces}
\parskip 0pt

\def\ctrline{\centerline}
\def\({\lbrack}
\def\){\rbrack}
\def\om{\omega}

\def\e{\ell}

\baselineskip 24pt

\def\aq{\lower 5pt\hbox to 50pt{\rightarrowfill}\atop
       {\hfill a\rarrow\scriptscriptstyle\infty\hfill}}
\def\rin{\lower 5pt\hbox to 50pt{{\rightarrowfill}
     \atop{\hfill r\rarrow\scriptscriptstyle\infty\hfill}}}

\rightline{WIS-93/48/Jun-PH}
\bigskip
\par
\def\tindent#1{\indent\llap{#1}\ignorespaces}
\def\refn{\par\hang\tindent}
\parskip 0pt
\centerline{\bf Generalized transparency in semi-inclusive processes}
\bigskip
\centerline {A.S. Rinat $^{\dagger}$ and B.K. Jennings}
\bigskip
\centerline {TRIUMF, 4004 Wesbrook Mall, Vancouver, Canada V6T 2A3}
\input tables
\vskip 1truein\noindent
\baselineskip 21pt
\par
{\bf Abstract:}

It is argued that the transparency of  a medium for passage of a nucleon,
knocked-out  in  a  semi-inclusive $(e,e'p)$  reaction  and  subsequently
scattered elastically, is not the same as  the one
measured in purely elastic
scattering.   Expressions   are  given   for  the   properly  generalized
transparency and  those are compared with  recently proposed, alternative
suggestions.   Numerical  results  are  presented  for  selected  nuclear
targets and  kinematic conditions, applying to  the Garino et al  and the
SLAC NE18 experiment.

\vskip.8truein\baselineskip 12pt

\vfil

$^\dagger$ Permanent address: Weizmann Institute of Science, Rehovot
76100,Israel
\eject\
\baselineskip 24pt

{\bf 1. Introduction}
\par

The  present  note  \footnote* {Protracted illness has caused
considerable delay in completing this ms. It supersedes a draft
which has been circulated well over a year ago.}
relates  to  recent  analyses   $^{1,2,3}$  of  some
semi-inclusive $A(e,e'p)X ^4, A(p,2p)X ^5$ processes
and attempts to extend the notion of
transparency of a medium as a measure of Final State
Interactions (FSI) effects of the knocked-out proton with the medium.
There is additional interest because of the predicted anomalously
large transparency if, part of the time,
the proton would appear as an object of small dimensions $^6$.

In the simplest  description of  semi-inclusive (SI) reactions (the Plane
Wave Impulse Approximation (PWIA)), the recoiling proton after
knock-out by the projectile exits without undergoing
FSI with the remaining core. Of course, scattering with medium particles
does occur and contributes to FSI.

An  accurate calculation  of the  latter under  general circumstances  is
close to impossible.   As an example of the  encumbering complications we
mention the fact that, in general, the knocked-out nucleon does not leave
the nuclear core in its initial state, i.e. the scattering may in part be
inelastic.  In addition the nucleon may
on its way out be excited and de-excited, or produce particles.  Some
of these processes have non-negligible amplitudes even for relatively low
energies.

It  is  natural  to  ascribe  the effects  of  elastic  scattering  of  a
knocked-out  particle  to  a  medium property  as  is  the  transparency.
However, it is  not at all evident how to link the latter to
the standard  definition of  transparency
in elastic scattering.   It thus seems
useful to  start with a  brief discussion  on the transparency  in purely
elastic  scattering,  then  to  evaluate   the  same  for  passage  of  a
knocked-out proton undergoing scattering and finally to compare the two.

The transparency of a medium probed in elastic scattering, of
say a proton, is defined as the probability
to find the latter traveling a distance $z$, thus
$$\Tscr^{el}(q,b,z)=|\psi_q^{(+)}(\vec r)|^2=
\left|e^{iqz+i\tilde\chi_q^{opt}(\vec r)}\right|^2\eqno(1.1)$$
Many-body dynamics provides an expression for the  above optical
(distortion)  phase. For instance, if the projectile momentum $q$ is
sufficiently larger than the average momentum $\langle p\rangle$
of a target  particle, the former
experiences the influence of essentially fixed
scattering centra. In those circumstances the above phase is conveniently
calculated in an eikonal  approximation, and in the  end averaged
over $\rho_A$, the normalized distribution probability of $A$ scatterers
$^7$ \footnote* {Note the difference between the above
densities and
those used for instance in Ref. 8: $\rho_A=(A!)^{-1}\rho_A'
;\rho_1=A^{-1}\rho_1'$, etc.}
$$\psi^{(+)}_q(\vec r)=e^{iqz}\left\langle e^{-i/v_q\sum_i 
\int_{-\infty}^z d\zeta V(\vec b_i-\vec b;\zeta-z_i)}
\right\rangle_{\rho_A},
\eqno(1.2)$$
where $v_q=E(q)/q\approx m/q$. The optical phase in (1.1) is determined
and subsequently approximated by
$$\eqalign{e^{i\tilde\chi_q^{opt}(\vec r)}
&\equiv \left\langle\prod_{i}e^{i\tilde\chi_q(\vec r,\vec r_i)}
\right\rangle_{\rho_A}
\approx{\rm exp}\left\(A\left\langle e^{i\tilde\chi_q(\vec r,{\vec r}_1)}
-1 \right\rangle_{\rho_1}\right\),}\eqno(1.3)$$
with
$$\tilde\chi_q(\vec r,\vec r_1)=\tilde\chi_q(\vec r-\vec r_1;-\infty,z)
=-v^{-1}_q\int_{-\infty}^z d\zeta
V(\vec b-\vec b_1;\zeta-z_1)=-v^{-1}_q\int_0^{\infty}d\zeta'
V(\vec b-\vec b_1;z-z_1-\zeta'),\eqno(1.4)$$
which is the eikonal phase due to the pair interaction
$V(\vec r-\vec r_1)$.

If the interaction $V$ has a range, small compared to nuclear
dimensions, the general off-shell 2-body eikonal phase $\tilde\chi
(\vec r;-\infty,z)$ may approximately be related to its on-shell
analogue $\chi(\vec b)\equiv\chi(\vec b;\infty,-\infty)$.
In a standard  parameterization
$$\eqalign{e^{i\tilde\chi_q(\vec r-\vec r_1;-\infty,z)}-1
&\to\theta(z-z_1)\left\(e^{i\chi_q(\vec b-\vec b_1)}-1\right\)\cr
e^{i\chi_q(b)}-1 &\approx
(2\pi/iq)f_q(0^{\circ})\int d\vec Q_{\perp}/(2\pi)^2
e^{i\vec Q_{\perp}\vec b}\(f_q(\vec Q_{\perp})/f_q(0^{\circ})\)\cr
&\approx (2\pi/iq)\delta^{(2)}(\vec b)f_q(0^{\circ})
\approx{1\o 2}\delta^{(2)}(\vec b)\sigma_q^{tot}(1-i\tau_q)}
\eqno(1.5)$$
For    given    momentum    $q$,   the    quantities    $f_q(Q_{\perp})$,
$f_q(0^{\circ}),\,\tau_q$ and $\sigma_q^{tot}$ in (1.5) are, respectively
the elastic scattering amplitude  for momentum transfer $\vec Q_{\perp}$,
the same  for $ \vec Q_{\perp} =0$, the  ratio of the real  and imaginary
part of the  latter and the total $NN$ cross  section. Incidentally,
Eqs. (1.3), (1.4)
(or the  parameterization (1.5)) enable a definition of $\Vscr^{opt}$,
which may be used instead of $\chi^{opt}$.

Returning to Eq. (1.1) one finally has
$$\eqalignno{\Tscr^{el}(q; b,z)&=e^{-A\sigma_q^{tot}\int_{z(b)}^z dz_1
\rho(b,z_1)}
\equiv e^{-\(z-\bar z(b)\)/\lambda} &   (1.6a)\cr
\Tscr^{el\,av}(q)&=\left\langle\Tscr^{el}(q,z)\right\rangle_{\rho}
={1\o {A \sigma_q^{tot}}}\int d\vec b\left\(
(1-e^{-At(b)\sigma_q^{tot}}\right\),
& (1.6b)\cr}$$
with  $\lambda$,  the  mean  free  path  and  $(\vec  b,\bar  z(b)$)  the
coordinate,  where  the projectile  enters  the  region of  non-vanishing
density.    Along  with   $\Tscr^{el}$,   one   defines  a   transparency
$\left\langle\Tscr^{el}(q)\right\rangle_{\rho}$,   averaged    over   the
density $\rho(\vec r)$, and Eq. (1.6b) is the standard result in terms of
the thickness function
$$t(b)=\int_{-\infty}^{\infty} dz\rho(b,z)\eqno(1.7)$$

For  later  reference  we  note  that the  densities  appearing  in  Eqs.
(1.3)-(1.7) are all diagonal.  In addition,  Eq. (1.6) does not contain a
pair-correlation function: Even if the  projectile is a nucleon it is
not part of the correlated target.

Consider next SI and TI reactions.
A convenient definition for a generalized transparency
 is $^4$
$$\Tscr^{reac}={{\rm Yield}^{exp}\o {{\rm Yield}^{PWIA}}}
\approx{{\rm Yield}^{th}\o {\rm Yield}^{{PWIA}}}
\eqno(1.8)$$
In semi-inclusive reactions  the detected proton is  knocked-out from the
nucleus and  not produced in  an accelerator. Consequently  the
above generalized  transparency cannot  without proof be  identified with
$\Tscr^{el\,av}$, Eq.  (1.6b)$^1$.  In fact, some  of the above-mentioned
analyses were  motivated by  an observation  on the  FSI of  the outgoing
proton in  the $(e,e'p)$  experiment of Ref. 4: when $\Tscr^{reac}\to
\Tscr^{el\,av}$, the  extracted mean-free
path  $\lambda$ exceeded  the  one  estimated from  the  free $NN$  cross
section in Eqs. (1.6a) by a factor of order of 2.

We thus focus in Section 2 on FSI in SI scattering. Assuming that high
momentum transfer SI cross sections factorize in the elementary one and
a response, we suggest the application of the non-perturbative, first cumulant
of the response$^{8,9}$, which describes FSI
generated by binary collisions $^{10}$. A basic assumption then enables the
derivation of a manageable expression for the SI response.
In particular for short-range $NN$ forces, a considerable simplification
results after parameterization in terms of total and reactive scattering
parameters.  A different derivation is given for an independent particle
model for target ground and excited core states and will be shown to come
close to the above result. With the thus derived FSI factors we give
expressions for generalized transparencies in SI processes. In
Section 3 we make a critical comparison of our approach
with the ones of Refs. 1-3. Numerical results are presented in Section 4.
In Section 5  we summarize our results and
mention their relevance for the delineation of
possible colour transparency effects in knock-out reactions.

{\bf 2. Transparencies in knock-out processes.}

Consider semi-inclusive $(e,e'p)$ scattering on a target with $A\gg 1$
when only the outgoing electron and the knocked-out proton are detected.
Let $\vec p$ be the momentum  of the  knocked-on particle and $\vec q$
the momentum transfer in the reaction. We consider the case,
when $\vec p'$, the momentum of the knocked-out particle
satisfies $p'=|\pq|\approx q\gg p$.
Assuming the $ep$ collision to take place on shell, the SI $(e,e'p)$
yield factorizes in the elastic $ep$ cross section and a generalized SI
response $S$ per nucleon, which is the quantity of our interest

$$\eqalign{S^{SI}(q,\omega,\vec p')&=1/A\sum_n
|\Fscr_{0n}(q)|^2\delta(\omega+\Delta_n-e(\pq))\cr
\Fscr_{0n}(q)&=\langle\Phi^0_A|\rho_q^{\dagger}|
\Psi^{(+)}_{n,\vec p+\vec q}\rangle\cr
\rho_q&=\sum_i{\rm e}^{i\vec  q\vec r_i}}\eqno(2.1)$$
In  Eq. (2.1)  above  we assumed  for simplicity  one  kind of  nucleons,
affected only by the longitudinal part of the $e-N$ interaction $\rho_q$.
The  squared   inelastic  form   factor  $\Fscr$  gives   the  transition
probability between the ground
state and scattering states, induced by the
above  density  fluctuation.  The  excited  states  are asymptotically  a
proton scattering  state with total on-shell  energy $e(\pq)$  and a
$A-1$ core in any conceivable excited state $n$, separated in energy from
the ground state by $\Delta_n$.

For sufficiently high energies only the incoherent part
of the density fluctuation matters, i.e. that part, where one and
the same particle (say particle $i$ in $\rho^{\dagger}_q,\rho_q$)
absorbs and emits the momentum transfer in the elementary collision
with the projectile.
Consider first the PWIA approximation  to (2.1) where one neglects
the interaction between the outgoing proton and the  core
$$V_i(\vec r_i)=\sum_{j \ne i} V(|\vec r_i-\vec r_j|)\eqno(2.2)$$
Its subsequent inclusion generates the FSI between the recoil proton
and the medium. Since an exact description  is generally impossible,
approximations cannot be avoided. Most use the high energy of
the recoiling proton. We first assume
factorization of scattering states in (2.1) in core states and a
proton wave function distorted by the core, i.e.
$$\Psi^{(+)}_{n,\vec p'}(\vec r_i;\vec r_j)
\approx \Phi^{A-1}_n(\vec r_j)
\psi_{\vec p'}^{(+)}(\vec r_i;\vec r_j)\eqno(2.3)$$
Eq. (2.3) implies that  the distortion of a fast proton in the field
$V_i$, Eq. (2.2), is the one by scatterers, fixed at positions $\vec r_j=
\vec b_j+z_j\hat q$.  As for
elastic scattering described by (1.3), the distorted wave of the
knocked-out proton
in the eikonal approximation is described by
$$\psi^{(+)}_{\vec p'}(\vec b_i,z_i;\vec r_j)={\rm exp}\(i(\pq)\vec r_j
+i\tilde\chi_q^A(\vec b_i;-\infty, z_i;\vec r_j)\),\eqno(2.4)$$
That distorted wave is then in a conventional fashion described by an
eikonal phase for given impact
parameter $\vec b_i$, accumulated on the path $(-\infty,z_i)\,^7$
$(v_{p'}=v_{\pq}\approx v_q)$
$$\tilde\chi_q^A(\vec b_i-\infty,z_i;\vec r_j)=-i/v_q\sum_{j\ne i}
\int_{-\infty}^{z_i}d\sigma V_i(\vec b_i-\vec b_j,\sigma-z_j)\eqno(2.5)$$

At this point we envisage the possibility of particle production and
absorption in $NN$ collisions. Although by choice not
detected in $(e,e'p)$ reactions, those particles could be present,
in which case the states (2.3) are insufficient and others
should be introduced explicitly. If absent in final states,
intermittent mesons may still be produced and absorbed in intermediate
stages. Those may be accounted for, using a complex
optical phase in Eqs. (1.1) and (1.3) or, alternatively, a
complex $NN$ potential $V\,\,^3$,
effectively describing elastic scattering, without the need to
specify inelastic channels containing mesons.

In the actual construction of the SI response we shall
follow two routes. One method employs
closure, while in the second one exploits the advantages of a
single particle model for all states in (2.1).
\par
{\bf 2a. Closure in the construction of the semi-inclusive response.}
\par
Assume that the energy transfer $\omega$ is larger than all
excited energies which contribute
significantly to the inelastic form factors in (2.1). One may then apply
closure  over  core  states, replacing the separation energies there by
an appropriate average $\langle\Delta\rangle$.\footnote*
{Whenever not leading to confusion we shall replace
$\vec r_k\to k;\,j(=\vec r_j)$ denotes all coordinates except the
one which is singled out: $i$ in the case above.}
Putting without impunity
$i\to 1$, the response (2.1) can be recast into the form (recall
$\vec p'=\vec p+\vec q)$
$$\eqalign{S^{SI}(q,y_0,\vec p')&=(m/q)\delta(y_0-p_z)\int d\vec r_1
\int d\vec r_1'\,\rho_1(1,1'){\rm exp}\(i\vec p.(1-1'\)
\(\Pi_{j\ge 2}\int\,dj\)\(\rho_A(1,1';j)/\rho_1(1,1')\)\cr
&{\rm exp}\left\(i\tilde\chi_q^A(\vec b_1;-\infty,z_1;j)
-i\tilde\chi_q^{*A}(\vec b_1';-\infty,z_1';j)\right\),}\eqno(2.6)$$
where the energy transfer $\om$ has been replaced by
the PWIA scaling variable$^{11}$
$$y_0=-q+\{2m(\omega-\langle\Delta\rangle)\}^{1/2}\eqno(2.7)$$
Again we meet in (2.6) $\rho_A$, a $A$-particle  density matrix, but
now diagonal in all coordinates $j$, except the one of the struck
nucleon. This reflects
the  single-particle nature  of  the
excitation operator:  Inclusion of  multi-particle exchange  densities or
currents,  would  lead  to   generalized  densities,  non-diagonal  in  a
corresponding number of coordinates.

Since the definition (1.8) of transparency in a knock-out reaction
relates to the response without FSI we start with  the PWIA
when $V,\chi \to 0$ in (2.5), thus
$$S^{SI,PWIA}(q,y_0,\vec p')\approx{m\over q}\delta(p_z-y_0)n(p)
\eqno(2.8)$$
Returning to Eq. (2.6) for a system where the recoiling proton
interacts with the core, the exact SI response may be written as
$$S^{SI}(q,y_0,\vec p')=(m/q)\delta(y_0-p_z)
\int\int d\vec r_1 d\vec r_1' e^{i\vec p(\vec r_1-\vec r_1')}
\rho_1(\vec r_1,\vec r_1')
\tilde R^{SI}(q,\vec r_1,\vec r_1')\eqno(2.9)$$
For homogeneous  matter the non-diagonal single-particle density in Eq.
(2.9) is a function of the difference of the arguments, and its Fourier
transform is simply related to the single particle momentum distribution
$${\rho_1(s,0)\o {\rho}}=\int\p3 e^{i\vec p\vec s}\,n(p)\eqno(2.10)$$
Although well-defined, (see for instance Section 2b below) no
simple expression exists for $\rho_1(\vec r_,\vec r_1')$ for
finite systems. A suitable interpolation is $^{3,12}$ ($\vec S=(\vec r_1+
\vec r_1')/2);\,\vec s=\vec r_1-\vec r_1';\rho_1(r,r)\to\rho(r))$
$$\eqalign{\rho_1(\vec r_1,\vec r_1')&\approx\rho(\vec S)\int d\vec S'
\rho_1(\vec s,\vec S')=\rho(S)\Sigma(s)\cr
\Sigma(s)&=\int\p3 n(p)e^{-i\vec p\vec s},}\eqno(2.11)$$
Assuming (for large nuclei) in (2.11) $\rho(S)\to\rho(r_1)$, substitution
into  Eq. (2.9) leads to
$$\eqalign{S^{SI}(q,y_0,\vec p')&=(m/q)\delta(y_0-p_z)\int d\vec r_1
\rho(r_1)\int d^3p"/2\pi)^3\,n(\vec p-\vec p")
R(q,\vec r_1,\vec p")\cr
&=\int d\vec r_1\rho(r_1)\int d^3p"/(2\pi)^3
S^{SI,PWIA}(q,y_0,\vec p'-\vec p")
R(q,\vec r_1,\vec p"),}\eqno(2.9')$$
where $R$ is the Fourier transform of $\tilde R$ in the variable $s$.

We notice that in both (2.8) and (2.9) (or (2.9$'$))
the assumption of an average separation energy (or closure) has caused
the replacement of the standard single nucleon spectral function$^{13}$
by the momentum distribution, as if the spectral strength is concentrated
in one energy. In contradistinction,
the second line in (2.9$'$) holds independent of
that assumption. In the ultimate expression for the transparency,
Eq. (1.8) one needs the $ratio$ of both expressions. It is plausible
that that ratio is less sensitive to the closure assumption than
the same for numerator and denominator separately. Next we observe that
the SI response contains a convolution and not a
product of the momentum distribution and the FSI factor. Below we shall
discuss examples and exceptions to general behaviour.

{}From a comparison with (2.8) one sees that
all FSI effects appear concentrated in the factor
$\tilde R^{SI}(q,\vec r_1,\vec r_1')$. The latter is far too complicated
for an exact evaluation. As above for elastic scattering (cf. Eqs. (1.2),
(1.3)), we exploit also here the high energy of the knocked-out nucleon
and assume that binary collisions of that nucleon with partners in the
core dominate the FSI in the SI response.
In the appropriate first cumulant approximation to $R$ one can perform
the integrals over $A-3$ coordinates in (2.6). This permits a reduction
of the $A$-body density matrix in (2.6) to
a (off-diagonal) 2-particle density matrix, which
upon introduction of an off-diagonal 2-body correlation function  can
be written as $^{8,9}$
$$\eqalign{\rho_2(1,2;1',2)&=\rho(2)\rho_1(1,1')\zeta_2(1,1';2)\cr
\zeta_2(1,1';2)&\approx \sqrt{g(1,2)g(1',2)\,},}\eqno(2.12)$$
where in the second line we give an
interpolation, suggested by
Gersch et al$^9$ and which shall be used throughout.
Notice the differences with the elastic case discussed in Section 1
and the relevant remark there after Eq. (1.7): In a SI process the
proton to be knocked-out is part of the correlated target, hence the
appearance of a correlation function. In contradistinction to the elastic
case, an essentially
different mechanism described above causes here densities and
generalized correlation functions to be non-diagonal.

Consider now Eq. (2.6) in the first cumulant approximation $^{10}$ and
concentrate on the the approximation $\vec b_1'=\vec b_1$. Using (2.5)
and (2.12) one finds ($\vec r=\vec r_1-\vec r_2$)
$$\eqalign{&\(\Pi_{j \ge 2}\int dj\)
\lbrace\rho_A(1,1';j)/\rho_1(1,1')\rbrace
{\rm exp}\left\(i\tilde\chi_q^A(\vec b_1;z_1,-\infty;\vec r_j)
-i\tilde\chi^{A*}_q(\vec b_1;z'_1,-\infty;\vec r_j)\right\)\cr
&=\left\(\Pi_{j \ge 2}\int dj
\lbrace\rho_A(1,1';j)/\rho_1(1,1')\rbrace
{\rm exp}\(i\tilde\chi_q^A(\vec b_1-\vec b_j;z_1,-\infty;z_j)
-i\tilde\chi^{A*}_q(\vec b_1-\vec b_j;z'_1,-\infty;z_j)\)\right\)\cr
&\approx 1+(A-1)\int d2{\rho_2(1,1';2)\o {\rho_1(1,1')}}
{\rm exp}\left\lbrace -i/v_q\int_{z_1'}^{z_1}d\sigma
V(\vec b_,\sigma-z_2)-2/v_q\int_{-\infty}^{z_1'}d\sigma
{\rm Im} V(\vec b,\sigma-z_2)-1\right\rbrace\cr
&\approx {\rm exp}\left\((A-1)\int d2 {\rho_2(1,1';2)\o
{\rho_1(1,1')}}{\rm exp}{\left\lbrace -i/v_q\int_0^{s_z}d\sigma'
V(\vec b,z-\sigma')-2/v_q\int^{\infty}_{s_z}d\sigma'
{\rm Im} V(\vec b,z-\sigma')-1\right\rbrace}\right\)}\eqno(2.13)$$

We shall use the parameterization (1.5) for short-range interactions in
the limit $Q_0\to\infty$ and shall assume that, as for nuclear matter,
$\zeta_2(\vec r_1,\vec r_2, \vec s)\approx \zeta_2(\vec r,s_z)$. One may
then rewrite the SI response Eq. (2.9), containing
binary collision FSI effects in the manageable form
$$\eqalignno{S^{SI}(q,y_0,\vec p')&=(m/q)\delta(y_0-p_z)
\int\int d\vec r_1 d\vec s e^{i\vec p\vec s}
\rho_1(\vec r_1,\vec s)\tilde R^{SI}(q,\vec r_1,\vec s)\cr
\tilde R^{SI}&=\tilde R^{tot}\tilde R^{reac}\cr
\tilde R^{tot}(q,\vec r_1,\vec s)
&\approx{\rm exp}\left\(-(A-1)(\sigma^{tot}_q/2)
(1-i\tau_q)\int_0^{s_z}dz
\rho(b_1,z_1-z)\zeta_2(0,z,\vec s)\right\)   &(2.14a)\cr
\tilde R^{reac}(q,\vec r_1,\vec s)
&\approx{\rm exp}\left\(-(A-1)\sigma^{reac}_q
\int^{\infty}_{s_z}dz
\rho(b_1,z_1-z)\zeta_2(0,z,\vec s)\right\)& (2.14b) \cr} $$
Eqs. (2.14) feature the total $NN$ total cross section, as well as the
'partial' and total $reaction$ cross sections
$$\eqalign{\sigma_q^{p,reac}(b)&\equiv 1-e^{-2{\rm Im}\chi_q(b)}
\approx \delta^{(2)}(\vec b)\sigma_q^{reac}\cr
\sigma_q^{reac}&=\sigma_q^{tot}-\sigma_q^{el}}\eqno(2.15)$$
We note that the first part in the exponent of $\tilde R^{tot}$ in Eq. (2.14a)
exists for real as well as for complex $V$ (or $\chi$) and decreases with
increasing $NN\, total$ cross section.  The
absorptive part of $\chi$, produces a component
(2.14b) in $\tilde R$ and $\Tscr$, which  decreases with the  $NN$
(total) $reaction \,(inelastic)$ cross  section.

We shall return to the results of Section 2a, but
elaborate first on the above mentioned alternative approach.
\par
{\bf 2b. An extreme single particle model}
\par

Using (2.3) we consider the part $\Fscr^i$
of the inelastic form factor
in (2.1) contributing to the response (2.1), which is excited by the
density fluctuation of the $i^{th}$ particle
$$\Fscr^i_{0n}(q)=\langle\Phi^A_0(i;j)|e^{-i\vec q\vec r_i}|
\Phi^{A-1}_n(j)\psi^{(+)}_{n,\pq}(i;j)\rangle\eqno(2.16)$$
For its evaluation we formulate the following single
particle model. If the knock-out is fast, i.e. if rearrangement can
be neglected, the orbital of the knocked-out particle $\phi_{\nu_i}(i)$
also determines the excited state of the remaining $A-1$
nucleons. One then finds for the product of
the two bound state wave functions in (2.16)$^{14}$
$$\eqalign{\Phi^A_0(i;j)\Phi_{\nu_i}^{A-1}(j)
&\approx\phi_{\nu_i}(i)\rho^{\nu_i}_{A-1}(j)\approx
\phi_{\nu_i}(i)\rho_{A-1}(j)\cr
\rho_{A-1}(j)&=\int di\,\,\rho_A(i;j) }\eqno(2.17)$$
It has been
assumed above that various $A-1$ particle diagonal densities
for states which differ from the ground state by one particle in orbital
$\nu_i$, may all be replaced by the density $\rho_{A-1}\equiv
\rho^0_{A-1}$ in the ground state. Of course, in order for Eq. (2.17) to
hold for every $i$, one must have an extreme single-particle model
for both target and core states.

Next we treat again the many-body eikonal phase in (1.2) in the first
cumulant (1.3) and the short-range approximation (1.5), and obtain for
the contribution of the $i^{th}$ particle to the inelastic form factor
(2.16)
$$\langle\Phi_A^0|e^{-i\vec q\vec r_i}|
\Psi^{(+)}_{\nu_i,\vec p'}\rangle
\approx\int d\vec r_i\phi_{\nu_i}(\vec r_i)e^{i\vec p\vec r_i}{\rm exp}
\left\(-(A-1)(1-i\tau_q)\sigma_q^{tot}/2\int^{z_i}_{-\infty}dz_2
\rho(b_i,z_2)\right\)\eqno(2.18)$$
Substitution in Eq. (2.1) gives
$$\eqalign{ S^{SI}(q,\om,\vec p')
&\approx A^{-1}\sum_{\nu}\delta(\om+\Delta_{\nu}-e_0(\pq))
\int\int d\vec r_1 d\vec r_1'
\phi^*_{\nu}(\vec r_1)\phi_{\nu}(\vec r_1')
e^{i\vec p(\vec r_1-\vec r_1')}\cr
&{\rm exp}\left\(-(A-1)\sigma^{tot}_q/2
\left\lbrace(1-i\tau_q)\int^{z_1}_{-\infty} dz_2\rho(\vec b_1,z_2)
+(1+i\tau_q)\int^{z_i'}_{-\infty}dz_2
\rho(\vec b'_1,z_2)\right\rbrace\right\)}\eqno(2.19)$$
In order to enable a comparison with the approach in Section 2a we
replace also here the separation energies by an average $\langle
\Delta\rangle$. The remaining sum over $\nu$ in (2.19) is then just the
expansion of the non-diagonal density matrix $\rho_1^{sp}
(\vec r_1,\vec r_1')$ in the extreme single particle model, i.e. with
orbitals which have unit occupation probability
$$\eqalign{ S^{SI}(q,y_0,\vec p')
&\approx (m/q)\delta(p_z-y_0)\int\int d\vec r_1 d\vec r_1'
\rho_1^{sp}(\vec r_1,\vec r_1')
e^{i\vec p(\vec r_1-\vec r_1')}\cr
&{\rm exp}\left\(-(A-1)\sigma^{tot}_q/2
\left\lbrace(1-i\tau_q)\int^{z_1}_{-\infty} dz_2\rho(\vec b_1,z_2)
+(1+i\tau_q)\int^{z_i'}_{-\infty}dz_2
\rho(\vec b'_1,z_2)\right\rbrace\right\)}\eqno(2.20)$$
Finally taking also here $\vec b_1'=\vec b_1$ (cf. (2.9))
$$\eqalign{S^{SI}(q,y_0,\vec p')&\approx (m/q)\delta(p_z-y_0)
\int\int d\vec r_1 d\vec r'_1\rho_1^{sp}(\vec r_1,\vec r'_1)
e^{i\vec p(\vec r_1-\vec r_1)}\cr
&{\rm exp}\left\(-(A-1)\sigma^{tot}_q/2
\left\lbrace\int^{z_1'}_{-\infty} dz_2 \rho(b_1,z_2)+(1-i\tau_q)
\int^{z_1}_{z'_1}dz_2\rho_1(b_1',z_2)\right\rbrace\right\)\cr
&\tilde R(q,\vec r_1,\vec s)={\rm exp}\left\(-(A-1)\sigma^{tot}_q/2
\left\((1-i\tau_q)\int^{s_z}_0 dz\rho(b_1,z_1-z)+2\int^{\infty}_{s_z}dz
\rho(b_1,z_1-z)\right\)\right\)\cr
&\rho_1^{sp}(\vec r_1,\vec r_1')=\sum_{\nu}\phi_{\nu}^{*}(\vec r_1)
\phi_{\nu}(\vec r_1'),}\eqno(2.21)$$
Eq.  (2.21)  resembles its  counterpart  Eqs.  (2.14) of  section  2a.
Based on a strict single particle model, it manifestly lacks reference to
pair  correlations, non-diagonal or diagonal. Otherwise it has the total
reaction cross section in (2.14b) replaced by the total $NN$  cross
section. The lack of correlations is understandable since we are using an
extreme single particle model.

The  other difference  with Eq. (2.13b), namely the
replacement of the reaction cross-section by $\sigma^{tot}$  is
more subtle. The assumption (2.17) prescribes core excited states to be
the target ground state from which one particle is removed. No other
excited core states are permitted, whereas in the closure approximation
of Section 2, particle knock-out from the core or general break-up
is included. In experiments with sufficient energy resolution to
guarantee that only  one nucleon is knocked out of the nucleus,
one expects the appearance of the total, rather than of
reaction cross-section. This may for instance
be the case for the NE18 SLAC experiment$^{15}$.  In
$'$less$'$ exclusive reaction  the partial use of  the reaction
cross-section may well  be more appropriate.

\par
{\bf 2c. The generalized transparency in SI reactions.}
\par

We are now in a position to give expressions for the desired SI
transparency (1.8)
$$\Tscr^{SI}(q,\vec p')\equiv {d\sigma (q,\omega,\vec p')^{data}\over
d\sigma(q,\omega,\vec p')^{PWIA}}
={S(q,\omega,\vec p')^{SI\,data}\over S(q,\omega,\vec p')^{SI\,PWIA}}
\eqno(2.22)$$
and in particular when in the FSI factor (cf. (2.12)) only
binary collision approximation are retained
$$\eqalignno{
\Tscr(q,\vec p')&={\int\int d1\,d1'\rho_1(1,1')e^{i\vec p(1-1')}
\tilde R(q,1,1')\o {\int\int d1\,d1'\rho_1(1,1')e^{i\vec p(1-1')}}}
&  (2.23a)\cr
&\approx \(n(p)^{-1}\)\int d1\rho(1)
\int d\vec p"/(2\pi)^3
n(\vec p-\vec p") R(q,1,\vec p")& (2.23b)\cr}$$
As had already been observed after Eq. (2.9$'$), the SI transparency (the
same holds for  the TI response) is generally expressed  as a convolution
in  momentum  space.   A  common  product in  (2.23a)  would  reduce  the
generalized  transparency to  the  appropriate FSI  factor  $R$ and  this
indeed happens  in some simplified models  (see Section 3). Note
that closure apparently renders $\Tscr$ independent of $\omega$.

Eqs.  (2.23)  will  be  the  basis  for  actual  numerical  calculations.
We remark that,
in principle, we could have exploited the Fourier transform
of  the FSI  factor Eq.  (2.9)  before application  of the  short-range
approximation (1.5).  It  is the most general form in  the first cumulant
approximation, and  the corresponding FSI factor  (or transparency) there
include,  in  principle  calculable off-shell  $NN$  scattering  effects.
Those are  lost in  the measurable, on-shell  expressions in  Eq. (2.14),
which are much  easier  to  handle, give  direct  insight in  parametric
dependence and also  happen to be most convenient  for making comparisons
with other  approaches.  An  example with the  result (2.23)  has already
been discussed  in the previous  subsection: Others will be  discussed in
Section 3.

\par
{\bf 3. Comparison}
\par
The reasoning  which, through  the parameterization  (1.5)
transforms the FSI  factor in  Eqs. (2.6),  (2.13) to  the simple  forms
(2.14) and (2.23), has in part been suggested  before by Kohama, Yazaki
and Seki (KYS). It is thus appropriate to
make first a comparison with
their work. In a first and inessential assumption, the $A$-particle
density in (2.6)  is taken by KYS in a  mean field approximation without
correlations (cf. Section 2b) $^3$
$$\rho_A(1,1';j)=\rho_1(1,1')\Pi_{j\ge 2}\rho(j),\eqno(3.1)$$
with the Negele-Vautherin Ansatz (2.11) for
the non-diagonal single  particle density $\rho_1(1,1')$. In Ref. 3b
a non-diagonal pair-distribution function (2.11) is used, as had
been suggested by Gersch et al$^9$.

Also KYS have to address the intractable eikonal phases in (2.6). Their
approximation centers on the oscillating exponent in (2.9) which KYS
claim to vary much faster than any other function in the integrand
of (2.6), thus permitting one to put
$\vec s=\vec r_1-\vec r_1'=0$ everywhere. An exception
is made for the non-diagonal single particle density (2.11), thereby
recognizing the basic role of that density for SI reactions.
When the above substitution is actually applied in (2.14)
one obtains for the SI response (2.6)
$$S^{SI\,KYS}(q,y_0,p')={m\o q}\delta(y_0-p_z)\int d\vec s
e^{i\vec p\vec s}\Sigma(s) \int d\vec r_1\rho(r_1)
e^{-(A-1)\sigma_q^{reac}\int^{z_1}_{-\infty}d\bar z\rho(b,\bar z)}
\eqno(3.2)$$
The above crucial assumption renders $\tilde R$ independent of $s_z$
and explicitly proportional to $n(p)$ and produces
a transparency
$$\Tscr^{SI\,KYS}(q)=\(A-1)\sigma_q^{reac}\)^{-1}\int d\vec b
\(1-e^{-(A-1)\sigma_q^{reac}t(b)}\)\eqno(3.3)$$
Eq. (3.3)  is not a convolution  as the general result  (2.23b) predicts.
Possibly  anticipating  (3.3),  Kohama  et  al,  in  fact,  $define$  the
transparency by (2.23b), but  $assume$ the involved ratio to
be  independent  of  the  knocked-out   proton  momentum  $p$  which,  as
demonstrated, is not generally the case.

Eq. (3.3) for the SI response in the KYS
approximation has obvious drawbacks, in part related to the fact
that the FSI factor Eq. (3.3) contains the reaction cross section.
For  energy  losses  below  the production threshold,
$\sigma^{reac}(q)=0$  and the manifest absence of an 'elastic' FSI
contribution (2.14a) leads to $R=1$, which is clearly incorrect. Since
well above
that threshold $NN$ reaction and total cross sections are of the same
order, it is likely that the above shortcoming is significant in general.

In addition $\Tscr^{SI\,KYS}$, Eq. (3.3) is independent of $p$, the
momentum of the outgoing nucleon. Thus, in spite of the fact that KYS
retained the non-diagonal single particle density matrix and obtained
a non-trivial SI response (3.2), the corresponding transparency
(3.3) is barely different from the elastic transparency (1.6b)  and
requires only the replacement of
the reaction by the total cross section.
The simple cause of all points raised is the exponent in (2.18). In fact,
since $p<p_F,\,\,p|\vec r_1-\vec r_1'|=\Oscr(1)$, its oscillations
are not conspicuously fast and consequently there
is no compelling  reason to  put  $\vec  r_1'=\vec r_1$. The
assumption  $\vec b_1'=\vec b_1$, but $z_1'\neq z_1$ applied
to (2.13) is apparently a much weaker one and
gives radically different results. In particular it
leaves in (2.14a) below the lowest inelastic threshold
an elastic part which is missing in (3.3).

Next we cite an expression for the transparency
given by Pandharipande and Pieper $^1$
$$\eqalign{\Tscr^{SI,PP}(q)&=(A-1)^{-1}\int d\vec b dz'\rho(b,z'){\rm
exp}\(-\int_{z'}^{\infty}
d\bar z\sigma_q^{tot}\big(\rho(b,\bar z)\big)g\rho(b,\bar z)\)\cr
&\approx\((A-1)\langle g\rangle \langle\sigma^{tot}_q\rangle\)^{-1}
\(\int d^2\vec b (1-e^{-(A-1)<\sigma^{tot}_q><g>t(b)}\)}\eqno(3.4)$$
The steps used to produce (3.4) again lead to a transparency independent
of the out-going proton momentum and which is similar to (1.6b) for
$elastic$ scattering (See Ref. 16 for a derivation emphasizing the above).

{}From the above it should have become clear that there are basic
differences between the transparency of a medium for an elastically
scattered proton and one which in a reaction is removed
from the target before scattering.
In the $'$elastic$'$ case the proton is not correlated at all with
nucleons in the nucleus. The interaction with the latter is through
the $diagonal$ density and this remains the case for the cross section:
Indeed $'$elastic$'$ transparency is a classical concept.

In contradistinction, the emerging proton before knock-out in either
a SI reaction, is part of the target. It is therefore
correlated with a core nucleon it interacts with, and the
nature of the process causes the densities in the cross section to
be non-diagonal in a non-trivial manner. No heuristic
reasoning can bridge  these intrinsically different pictures.
\footnote*{
It is  argued in  Ref. 1  that a pair-distribution  function in  (3.4) is
needed  to account  for  modifications  of the  particle  density due  to
short-range  correlations.   A  self-consistent   theory  of  the  target
accounts treats correlations in derived observables and for instance
the density depends implicitly on them.}

Next we compare the approach of the authors in Refs. 1 and 3.  Indeed, in
spite of the similarities, Eqs.  (3.3) and (3.4) are basically different.
KYS  realized the essential  appearance  of  non-diagonal densities:  The
resemblance  of (3.3)  to  (3.4)  is the  result  of the  above-discussed
specific approximation.

In concluding this section we  recall that, disregarding differences, the
SI transparencies  (3.3) and (3.4)  result if  in (2.23a), based  on Eqs.
(2.14) or  (2.21), $s_z\to 0$.  On  may then ask whether  the FSI factors
$\tilde R$ peak at $s_z=0$.  An  estimate can be made if the density
and $\zeta$  are replaced  by averages, which  results in  a $z$-integral
$\propto  s_z$.   The  FSI   factor  thus  decreases  exponentially  from
$s_z\approx0$, but the maximum is hardly pronounced $^{17}$.

\par
{\bf 4. Numerical results.}
\par

In the following  we report on results for the  transparency $\Tscr$, Eq.
(2.23), in its  dependence on the one-body  density matrix $\rho_1(1,1')$
and the semi-inclusive FSI factor $\tilde R(q,1,1')$.  Two versions have
been tested, namely the  closure method  (CL) based  on  Eqs. (2.11)  and
(2.14), and the extreme single particle model version (SP), using (2.21).

Input elements  in the  former are  total and  total reactive  $pN$ cross
sections, the  ratio's $\tau$ of real  and imaginary part of  the forward
elastic $NN$ amplitude in the  parametrization (1.5) and the non-diagonal
pair distribution  function $\zeta_2$  in the approximation  (2.12).  The
required density  matrix has been  computed from (2.10) and  (2.11), with
single particle  densities and momentum  distributions as in Ref.  3.  In
the SP model the density matrix $\rho_1^{sp}$, required in (2.21)
is constructed from single-particle  wave functions, chosen to correspond
to levels  in a Saxon-Woods  potential and  which are occupied  with unit
probability till a given $A(Z,N)$ is reached.

Predictions have  been made for  two sets of kinematic  conditions.  We
considered  first the  experiment  of  Garino $et\,  al$  on the  targets
$^{12}$C, $^{27}$Al, $^{58}$Ni and $^{181}$Ta$^4$.  The momentum transfer
was  approximately  constant,  $q\approx  610$ MeV,  while  the  outgoing
protons had energies $E_{p'}=180\pm 50$ MeV.   We refer to Ref. 4 for the
extraction  method  for $\Tscr^{SI}$,  Eq.  (2.22),  from rather coarsely
binned data.

Calculations have been  made for varying $p_z$,  with average $p_z\approx
0$;  small $p_{\perp}$  have been  neglected.  Fig.  1 contains  data and
predictions for  the transparency $\Tscr$, Eq.  (2.23), for CL and  SP at
$p_z=0$.  Drawn, dashed and dot-dashed  curves correspond to CL and refer
respectively, to results based on  (2.14) with and without (off-diagonal)
pair correlations and for $s_z=0$,  correlations included.
The dotted curve is
the  SP  prediction (2.21). Compared to our  $'$best$'$
prediction, the  data appear underestimated  by 15-20$\%$.  Fig.  2 shows
for $^{58}$Ni the dependence of $\Tscr$ on the direction  of the
proton momentum, varying with respect to the momentum transfer. That
dependence is qualitatively the same for all targets.

The  second set  of conditions  corresponds  to the  SLAC NE18  $(e,e'p)$
experiment  on $^{12}$C,  $^{27}$Al, $^{63}$Cu  and $^{208}$Pb.   Table I
gives the much higher 3-momentum and  energy transfers, the square of the
4-momentum transfer  and the  momentum of the  outgoing proton,  which is
again close to $q$. Figs. 3a-d show predictions for $Q^2$=1.04, 3.00,
5.00, and 6.77 GeV: data have as yet not been released.  One observes:

i) Inclusion of correlations reduces  by roughly 15-20$\%$ the opacity of
each data point calculated with CL.

ii)  When compared  with CL  predictions under  comparable circumstances,
(i.e. with correlations included) the  KYS assumption  $s_z=0$
increases $\Tscr$ by 4-7 $\%$ for the kinematics of the Garino experiment
and $p_z=0$ , when $A$ increaces from 12 to 181.
Slightly smaller increases have been observed for NE18.

iii) A drastic reduction of  $\Tscr$ by a factor  of order
1.5-2.0, and in  addition a larger $A$-dependence results when comparing
SP and CL, both lacking correlations. Part of this reduction is due
to the sole appearance in (2.21)
of total cross sections, which always exceed total
reaction cross sections in (2.14).

iv) Only moderate  10$\%$ reductions occur in the  predicted $\Tscr$ when
$q$ in NE18 changes as much a factor 4.  Similar changes occur when going
from the Garino to the NE18 conditions and all can be predicted.
In a NR theory the relevant eikonal phases in (2.5)
are proportional to $v_q\propto m/q$,
which for relativistic kinematics are replaced by $E(q)/q$, which has a
much weaker $q$ dependence.
This is also  clear from the high-$q$ parametrization (1.5)
of the  on-shell eikonal phase  or amplitude: The $E(q)/q$  dependence of
the phase  $\chi$ cancels out in  the chain $v\to t\to\sigma$  and leaves
only  a moderate  implicit $q$-dependence  in cross  sections.  Otherwise
there are  for the NE18 kinematics  no conspicuous changes in  the trends
already  observed for  lower $q$  regarding $A$-dependence,  influence of
correlations and for the results of the KYS assumption.

Finally we report  on computations related to the  basic assumption $\vec
b_1'\to\vec b_1$ in (2.6), and  which through (2.13) enable evaluation of
the  SI response  in the  form (2.13).   It appears  difficult to  make a
reliable test.   We have  done so  approximately for  the SP  model using
harmonic oscillator wave  functions.  For those, we  calculated the ratio
of transparencies  under the  Garino conditions,  once for  $\vec b_1'\ne
\vec b_1$  , then with the  two equal.  From positive  to negative $p_z$,
the  above  ratio  decreases,  passing  1 in  the  neighbourhood  of  the
quasi-elastic peak, $ p_z \approx 0$.   At least there
the tested approximation seems to be fully justified.

Out  of the  alternative  approaches mentioned,  we  already discussed  a
comparison with KYS.  One would like  to do the same with the predictions
of  Ref. 1.   However, in  view of  the discussion  in Section  3 on  the
seizable differences in approach, such a step by step comparison with the
work  of  Pandharipande  and  Pieper$^1$  does  hardly  make  sense.   An
exception should  be made  for the effect  of correlations.   Although
disagreeing on the way $g(r)$ has been introduced in $\Tscr$,
a similar substantial increase in $\Tscr$ has, been
reported.   It  is of  interest  to  note  that (formally!)  a  different
$\zeta_2$ has been used in $^1$.  The same holds if the pair distribution
function for NM, available to us, is replaced by one for variable density
(cf.  Fig. 6,  Ref. 1).  This does not make  it likely that uncertainties
in the off-diagonal $\rho_2$ can  be blamed for the observed discrepancy.

Next  we come  to the  high $q,Q^2$  predictions by  Benhar $et\,al\,^2$.
Since based on Ref. 1, it is again difficult to make comparisons with our
results.  We only  note that for the Garino  experiment their predictions
exceed ours by 10-15$\%$ , but those  fall short by about the same amount
for high $q$.
\par
{\bf 5.  Summary and conclusion.} \par
\par
We addressed  above the transparency  of a  nuclear medium for  a nucleon
knocked-out  in semi-inclusive  reactions, defined  as the  ratio of  the
experimental  yield and  the same,  calculated  in the  PWIA and  clearly
related to  Final State Interaction  of the knocked-out nucleon  with the
core.  In  view of  the relatively high  momentum transfers  involved, we
used the first cumulant approximation  to the SI response, which focuses
on binary collisions  of  the  knocked-out proton  with  a core  nucleon.
Production  and absorption  of undetected  pions etc.,  are approximately
included, implying a complex interaction $V_{NN}$.  In
a  simple parameterization  for $V$,  ultimately both  the total  and the
total reaction (inelastic)  $NN$ cross section appear  in expressions for
the transparency.

We then  emphasized that  from the single-particle nature of the
excitation  mechanism of  an electron,
colliding with a single nucleon, it follows that the SI response or cross
section  is ultimately  proportional  to a  two-particle density  matrix,
non-diagonal  in the  coordinate  of the  struck  nucleon.  That  density
matrix can alternatively be expressed in terms of a $non-diagonal$ single
particle density and a pair-distribution function.

The above observation is basically  different for elastic scattering from
a medium, whether the projectile is  a nucleon or otherwise.  For it, the
cross section appears related to $diagonal$ densities.  As a consequence,
and notwithstanding intuitive reasoning, the transparency of a medium for
passing of a  proton removed in a reaction is basically dissimilar to the
same in elastic scattering.  In particular the correct expression depends
on the momentum of the outgoing proton.

Throughout  we  have treated  the  first  cumulant approximation  in  two
versions.   In the  first,  based  on closure  over  excited states,  one
explicitly   includes   (non-diagonal)  pair-correlations   between   the
knocked-out  proton and  nucleons in  the core.   The second  approach is
based on a  strict, single-particle model for the nuclear  ground state
as well as for excited states, and the treatment thus foregoes
pair-correlations effects.

Calculations have been performed for both versions using simple
parameterizations for the effective  $V_{NN}$, thereby avoiding the usual
encumbering off-shell  effects. The kinematic conditions considered are
those for the  experiments by Geesaman and Garino et al, and also
for  the high energy SLAC NE18 experiment.  For
the  former,   results  have  been   compared  with  the   above  defined
experimental quantities. The predictions are  10-15$\%$ in excess
of observed transparencies.

As   to  generalities   around  the   predictions,  we   obviously  found
transparencies decreasing with mass number and could show that the use of
parametrized, on-shell  eikonal scattering  phases produce  only moderate
energy   dependence   in   $\Tscr$.   The   inclusion   of   non-diagonal
pair-correlations leads to a sizable reduction of opacity.

As regards the two versions considered, the exclusive appearance of total
reaction cross sections  in the strict single particle  model in contrast
to total and (smaller)  total  reaction cross  sections  in the  closure
approach, causes an appreciable reduction in $\Tscr$ when compared to the
closure approach.

We then compared our approach with previously suggested ones.  Ours is in
spirit close  to  the one  by  Kohama, Yazaki  and  Seki who  also
emphasize  the   use  a  non-diagonal  single-particle   density  matrix.
Unfortunately a strong  assumption in the actual  evaluation, removes the
dependence of the SI transparency on  the momentum of the outgoing proton
and  in addition, renders  it solely  dependent on  the reaction  cross
section. When the latter is replaced by  the total cross  section, the
resulting SI transparency is indistinguishable from the elastic one.

We then discussed the approach of Pandharipande and Pieper, which differs
basically from ours. Their expression is essentially the elastic
transparency with
modifications we  discussed and criticized on the basis that the
projectile proton in elastic scattering  is essentially not correlated to
the  target particles,  whereas  the knocked-out  proton in  a  SI
reaction is.

At   this  point   we  only   briefly   mention  the   issue  of   colour
transparency$^{6}$, which envisages objects of smaller than hadronic size
to  be   produced.   Those  subsequently  exit   with  anomalously  large
transparency before  reverting to  their $'$normal$'$  hadronic identity.
For instance, Komara  $et\,al$ emphasize that in their  theory (cf.  Eqs.
(3.3) and (1.6b), reaction cross sections replace larger total $NN$ cross
sections, thus reducing SI  transparencies without invoking exotic colour
transparency of  sub-hadronic  constituents.  For  $(e,e'p)$  experiments
under  current conditions  the issue  appears  to be  academic: The  most
recent analysis of the dedicated SLAC NE18 ($e,e'p$) experiment for $Q^2$
up to 7GeV$\,^2$, shows no trace of colour
transparency$^{18}$.  The latter had already been called in$^2$, in order
to  explain  totally inclusive  electron  scattering  experiments in  the
$Q^2\approx  2-3$  GeV  region$^{19}$. In  on-going  and  planned
$(e,e'p)$  and  ($p,2p)$ reactions, the detection of colour  transparency
requires a maximally accurate theory for the regular, nucleonic
transparency, which in part motivated the study above.
\par
{\bf Acknowledgements}
\par
S.A. Gurvitz, J.P. Schiffer, R. Seki and  K. Yazaki are kindly thanked
for discussions and correspondence on the topic matter. B. Wirenga has
been kind enough to provide the pair distribution function for
nuclear matter.
\noindent

\par

{\bf References}
\bigskip

\refn{$^1$}
V.R. Pandharipande and S. C. Pieper, Phys. Rev. {\bf C45} (1992) 780

\refn{$^2$}
O. Benhar, A. Fabrocini, S. Fantoni, G.A. Miller, V.R. Pandharipande
and I. Sick, Phys. Rev. {\bf C44} (1991) 2328; Phys. Rev. Lett.
{\bf 69} (1992) 881

\refn{$^3$}
A. Kohama, K. Yazaki and R. Seki, Nucl. Phys. {\bf A536} (1992) 716;
$ibid$ {\bf A551} (1993) 687

\refn{$^4$}
D.F.  Geesaman $et\,al$, Phys Rev. Lett. {\bf 63} (1989) 734;
G. Garino $et\,al$, Phys Rev. {\bf C45} (1992) 780;

\refn{$^5$}
A.S. Carroll $et\,al$, Phys Rev. Lett. {\bf 61} (1988) 1698

\refn{$^6$}
E.g. G. R. Farrar, H. Liu, L. L. Frankfort and M. I. Strikman,
Phys. Rev. Lett. {\bf 61} (1988) 686

\refn{$^{7}$}
R.J. Glauber, $Lectures\, in\, Theoretical\, Physics$. Ed. W.E. Brittin
$et\, al$ (Interscience, NY., 1959) p. 315; Proceedings Int'l Conf. on
Particle and Nuclear Physics, Rehovot, Israel (1957) (Ed. G. Alexander,
NHPC)

\refn{$^8$}
See for instance A.S. Rinat and M.F. Taragin, Phys. Rev. {\bf B41}
(1990) 4247

\refn{$^9$}
H.A. Gersch, L.J. Rodriguez and Phil N. Smith, Phys. Rev. {\bf A5}
(1972) 1547

\refn{$^{10}$}
H.A. Gersch and  L.J. Rodriguez, Phys. Rev. {\bf A8} (1973) 905

\refn{$^{11}$}
I. Sick, D. Day and J.S. McCarthy, Phys Rev. Lett. {\bf 45} (1980) 871

\refn{$^{12}$}
J. Negele and D. Vautherin, Phys. Rev. {\bf C5} (1971) 1472; $ibid$
{\bf C11} (1974) 1031

\refn{$^{13}$}
E.g. A.E.L. Dieperink, T. deForest, I. Sick and R.A. Brandenburg, Phys.
Letters {\bf B63} (1976) 261; E. Pace and G. Salme, Phys. Lett.
{\bf B110} (1982) 411

\refn{$^{14}$}
B.K. Jennings and G. Miller, Phys. Rev. {\bf D44} (1992) 692

\refn{$^{15}$}
J. van den Brand $et\,al$, SLAC-NPAS proposal NE18 (1990); R.D. McKeown,
CalTech Preprint OAP-719(1992)

\refn{$^{16}$}
I. Mardor, Y. Mardor, E. Piasetzki, J. Alster and M. Sargsyan, Phys.
Rev. {\bf C46} (1992) 761

\refn{$^{17}$}
E.g. C. Carraro and A.S. Rinat, Phys. Rev. {\bf B45} (1992) 2945

\refn{$^{18}$}
P.W. Fillipone, TRIUMF Workshop on Colour Transparency, 6-9 Jan. 1993

\refn{$^{19}$}
D.B. Day et al, Phys. Rev. {\bf C40} (1989) 1011

{\bf  Figure captions}

\par\offset{55pt}{Fig. 1.} Transparencies for passage of knocked-out
protons in semi-inclusive $(e,e'p)$ reactions on different targets.
Data are from Ref 4. Drawn,
dashed and dot-dashed curves correspond to CL model based on Eq. (2.14)
and correspond respectively to calculations with and without (non-
diagonal) pair correlations, and $s_z=0$ (no correlations). The dotted
curve is for the SP model (2.14).

\par\offset{55pt}{Fig. 2} Dependence of the transparency of $^{58}$Ni
in its dependence on the direction of the
outgoing momentum with respect to the direction of the
momentum transfer.                                                       .

\par\offset{55pt}{Fig. 3a} The same as Fig. 1 for  the NE18 experiment
with $Q^2$=1.04 GeV (see Table I).

\par\offset{55pt}{Fig. 3b} The same as Fig. 3a: $Q^2$=3.00 GeV (see
Table I).

\par\offset{55pt}{Fig. 3c} The same as Fig. 3a: $Q^2$=5.00 GeV (see
Table I).

\par\offset{55pt}{Fig. 3d} The same as Fig. 3a: $Q^2$=6.77 GeV (see
Table I).
\bigskip
\baselineskip 14pt
\noindent
\vskip 2truecm
\ctrline{\bf Table I}
\medskip
 Kinematic conditions for SLAC NE18 experiment. In units
GeV$^{-1}$ the four columns give momentum, energy transfer, square of
4-momentum and momentum of outgoing proton.
\medskip
\begintable
$q$ | $\omega$ | $Q^2$ | $p'$    \crthick
1.198 | 0.630 | 1.038 | 1.201   \cr
2.439 | 1.719 | 2.995 | 2.454     \cr
3.538 | 2.742 | 4.997 | 3.543     \cr
4.484 | 3.652 | 6.772 | 4.490
\endtable
\end